\documentclass[prx,twocolumn,showpacs,amsmath,amssymb,superscriptaddress]{revtex4}
\usepackage{graphicx}
\usepackage{dcolumn}
\usepackage{bm}

\begin{document}


\preprint{prepared for Physical Review X}

\title{Experimental Observation of Bohr's Nonlinear Fluidic Surface Oscillation}

\author{Songky Moon}
\affiliation{School of Physics and Astronomy, Seoul National
University, Seoul 151-747, Korea}
\author{Younghoon Shin}
\affiliation{School of Physics and Astronomy, Seoul National
University, Seoul 151-747, Korea}
\author{Hojeong Kwak}
\affiliation{School of Physics and Astronomy, Seoul National University, Seoul 151-747, Korea}
\affiliation{Samsung Electronics Company, Hwaseong 445-330, Korea}
\author{Juhee Yang}
\affiliation{Russia Science Seoul, Korea Electrotechnology Research Institute, Seoul 151-747, Korea}
\author{Sang-Bum Lee}
\affiliation{Korea Research Institute of Standards and Science,
Daejon 305-340, Korea}
\author{Soyun Kim}
\affiliation{School of Physics and Astronomy, Seoul National
University, Seoul 151-747, Korea}
\author{Kyungwon An}
\email{kwan@phys.snu.ac.kr} \affiliation{School of Physics and
Astronomy, Seoul National University, Seoul 151-747, Korea}

\date{\today}

\begin{abstract}
Niels Bohr in the early stage of his career developed a nonlinear theory of fluidic surface oscillation in order to study surface tension of liquids.
His theory includes the nonlinear interaction between multipolar surface oscillation modes, surpassing the linear theory of Rayleigh and Lamb.
It predicts a specific normalized magnitude of $0.41\dot{6}\eta^2$ for an octapolar component,
nonlinearly induced by a quadrupolar one with a magnitude of $\eta$ much less than unity.
No experimental confirmation on this prediction has been reported. 
Nonetheless, accurate determination of multipolar components is important
as in optical fiber spinning, film blowing and recently in optofluidic microcavities for ray and wave chaos studies and photonics applications.
Here, we report experimental verification of his theory.
By using optical forward diffraction, we measured the cross-sectional boundary profiles at extreme positions of a surface-oscillating liquid column ejected from a deformed microscopic orifice.
We obtained a coefficient of $0.42\pm0.08$ consistently under various experimental conditions.
We also measured the resonance mode spectrum of a two-dimensional cavity formed by the cross-sectional segment of the liquid jet.
The observed spectra agree well with wave calculations assuming a coefficient of $0.415\pm0.010$.
Our measurements establish the first experimental observation of Bohr's hydrodynamic theory.
\end{abstract}


\maketitle

\section{Introduction}

\begin{figure}
\centering
\includegraphics[width=3.4in]{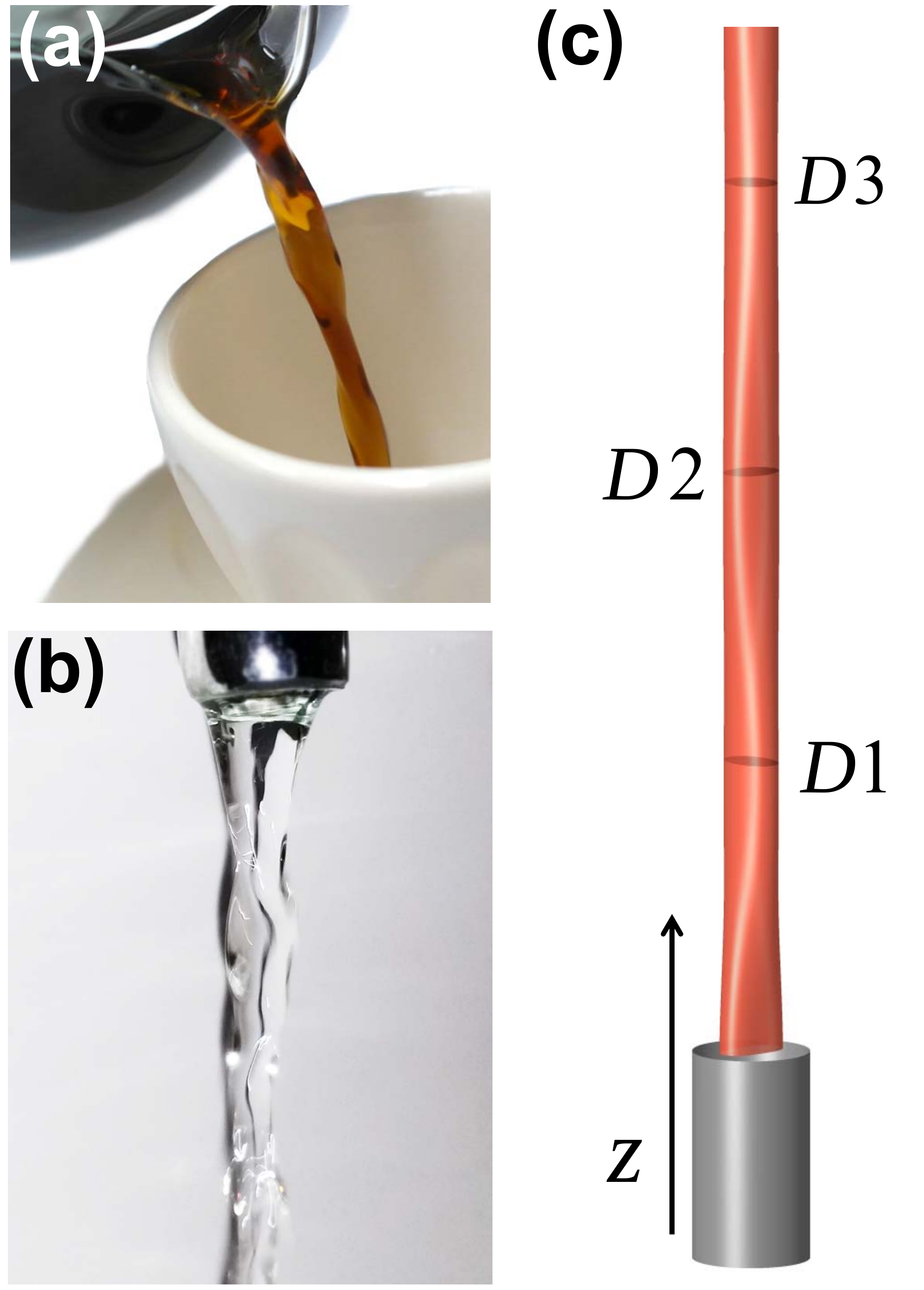}
\caption{Examples of the fluidic surface oscillation due to surface tension. (a) Coffee poured from a pot. 
(b) Water from a faucet. 
(c) A 3-D model of a vertically ejected liquid jet. The pattern is stationary while the liquid moves in $z$ direction. 
Extreme positions of surface oscillation with stationary cross sections are indicated as D1, D2, D3, and so on. 
}
\label{fig1}
\end{figure}

Optofluidics deals with a synthetic system of optical and fluidic elements \cite{Psaltis, Monat, Horowitz}, utilizing advantages of both areas \cite{Fainman}. As one of optofluidic examples, deformed microjet cavities are two-dimensional optical resonators supporting whispering-gallery-mode(WGM)-like cavity resonances \cite{Yang06}. A deformed microjet cavity is formed by a cross-sectional segment of a fluidic microjet, which undergoes surface oscillation as shown in Fig.\ \ref{fig1}. 
It provides not only a high quality factor owing to its clear and smooth surface but also high output directionality based on internal ray and wave dynamics. These features can be tuned in real time by controlling the flow rate or ejection pressure.
Deformed microjet cavities are known to be a versatile platform for studying quantum chaos, prototyping highly efficient light sources as well as various photonic devices \cite{Noeckel96, Lee07PRA, Lee07APL, Lee09PRA, Yang10}.

As is often the case with tunable optofluidic components, the optical characteristics of a deformed microjet cavity is sensitive to the detailed shape of its surface boundary. Even for non-optofluidic application such as optical fiber spinning and film blowing \cite{Bechtel}, accurate determination of multipolar components is required for consistent material processing. For optical cavity applications, in particular, it is necessary to measure the boundary profile of this cavity as accurately as possible. However, it is generally not easy to measure the accurate shape of a fluidic object. Liquid surface exhibits specular reflection and thus make the conventional optical triangulation technique inapplicable \cite{Narita}. Moreover, the continuous columnar shape of the microjet makes top-view imaging of its boundary simply impossible. 

It may seem that these difficulties in measurement might be easily compensated for by a theoretical approach. It is particularly because there exists a well-established hydrodynamic theory to describe the surface oscillation of a liquid jet, mostly owing to Lord Rayleigh and Sir Lamb \cite{Rayleigh78, Rayleigh79, Lamb}. 
However, this theory, based on the first-order approximation, cannot explain some detailed experimental facts from spectroscopic observations as to be discussed below. It was Niels Bohr at his early career who first extended the surface oscillation theory by including nonlinear interactions \cite{Bohr}. The extended theory and accompanying surface tension measurement of his own based on the lowest-order surface oscillation period enabled him to win a gold medal \cite{Bohr} in an academic contest of the royal Danish academy.  However, the surface oscillation period does not reveal the multipolar surface-oscillation amplitudes, which is the hallmark of Bohr's theory. So far, direct experimental verification of the multipolar surface oscillation amplitudes has not been reported.

In this paper, we present two independent experimental studies to determine the multipolar surface-oscillation amplitudes in the boundary profile of a deformed microjet cavity. One is a non-destructive surface reconstruction experiment based on optical forward diffraction. The other is the spectroscopy of the cavity resonances compared with wave calculation results for various cavity boundary profiles. We found the results of both studies exhibit good agreement with Bohr's prediction in nanometer-level accuracy. Our work thus marks the first experimental verification of Bohr's hydrodynamic theory. Moreover, precise knowledge of our microjet cavity boundary allows us to predict optical and dynamical properties of cavity resonances in advance. Quality factors, output directionality and intermode interactions can be predicted beforehand to guide actual experiments for various photonics applications.

This paper is organized as follows. In Sec.~\ref{sec2}, with the linear theory of Rayleigh and Lamb applied to our liquid jet, the basic properties of the jet are discussed. We then explain how optical cavity resonances are observed for our jet and discuss their inconsistency with the cross sectional shape of the jet predicted by the linear theory. The inconsistency is attributed to small shape perturbations, based on high sensitivity of intracavity ray dynamics and consequent cavity resonances on such small shape perturbations. In Sec.~\ref{sec4}, the principle of our optical forward diffraction technique and the experimental result on microjet surface profile measurement are presented. In addition, the wave calculation results assuming the surface profile obtained from the forward diffraction experiment are presented and shown to be consistent with the observed cavity resonances. In Sec.~\ref{sec5}, Bohr's nonlinear theory, its limitations and our extension are discussed. We then show Bohr's theory is adequate for our liquid jet under the current experimental condition. In Sec.~\ref{sec6}, we summarize our results with remarks on possible applications and future prospects.

\section{Application of linear analysis to a liquid jet} \label{sec2}

\subsection{A liquid jet to be analyzed}

A liquid stream from a non-circular orifice shows semi-periodic surface oscillations due to surface tension as a restoring force. Such oscillations are easily observed in our daily lives as illustrated in Fig.~\ref{fig1}. The deformed boundary shape of our microjet cavity is also due to this surface oscillation. In our setting, a liquid jet is vertically ejected from a near-elliptical orifice \cite{Yang06} as shown in Fig.~\ref{fig1}(c). When the ejection pressure is held fixed, the surface shape of the jet becomes stationary. The major and minor axes of the horizontal cross-sections are repeatedly exchanged as we move along the vertical axis. This phenomenon is known as `axis switching' \cite{Kasyap}. Moreover, the amplitude of surface oscillation decays and thus the cross-sectional shape would gradually converge to a circle.

\begin{figure}
\centering
\includegraphics[width=3.4in]{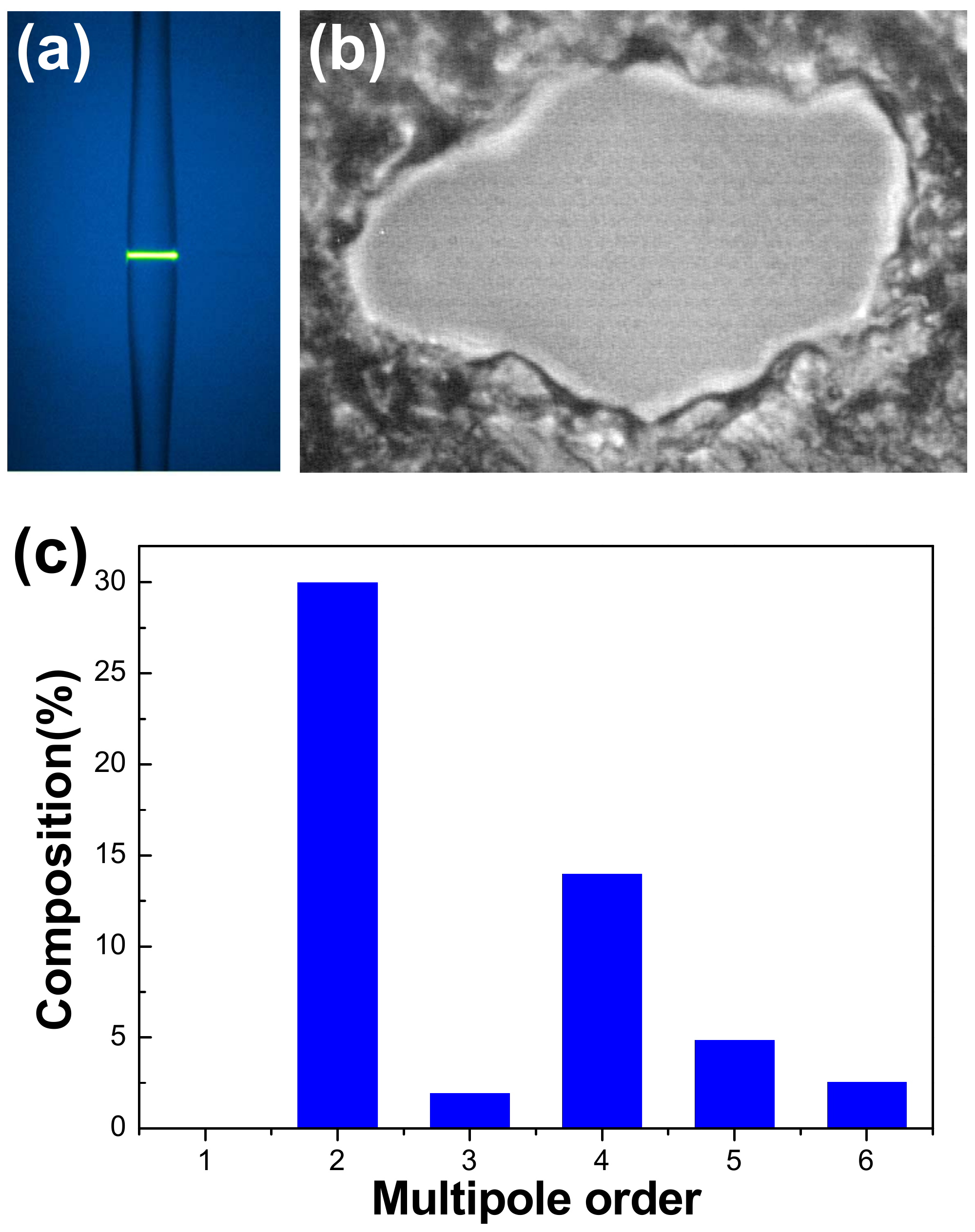}
\caption{Actual shape of our liquid microjet.
(a) Microscope image of the liquid jet column seen from the side. Mean radius is $15~{\mu}$m. 
The bright region of about 3 $\mu$m thickness is the segment excited by a pump laser beam to form a two-dimensional microcavity. 
(b) Microscope image of the orifice filled with liquid. 
(c) Multipolar decomposition of the boundary profile of the orifice of (b). 
The origin is chosen so as to make the dipolar component vanish. In this decomposition, quadrupolar (denoted by 2) and octapolar (denoted by 4) components are dominant. 
}
\label{fig2}
\end{figure}

A partial side view of the jet is shown in Fig.\ \ref{fig2}(a) and the microscope image of the orifice is shown in Fig.~\ref{fig2}(b). 
The shape of the orifice is decomposed into quadrupolar (labeled by 2) and octapolar (labeled by 4) components mostly. 
The decomposition serves as one of the initial conditions in the analysis below. 
From the microscope images of the jet like Fig.~\ref{fig2}(a), we can easily identify the directions of major and minor axes of cross-sectional shape. Moreover, the axis-switching is also clearly observed. The average period of the axis switching for the jet is $270 \pm10~{\mu}$m in Fig.~\ref{fig2}(a). 

\subsection{Linear Analysis}
 
According to the linear theory of Rayleigh and Lamb \cite{Rayleigh78, Rayleigh79,Lamb}, the $m$th mode of surface oscillation can be expressed in the cylindrical coordinates as
\begin{equation}
r_m=a \eta_m \cos(m \theta) \cos(k_m z + \xi_m) e^{-z/L_m}\;\;\;\; {(m=2, 3, \ldots)} \label{eq1}
\end{equation}
where $a$ is the mean radius, $\eta_m$ is the relative amplitude, ${k}_{m}\equiv 2\pi/\Lambda_m$ is the wave vector in $z$ direction with $\Lambda_m$ the oscillation wavelength, ${\xi}_m$ is the initial phase at the orifice and $L_m$ is the decay length.
For our liquid jet, $a$ is varied in the range of $11-15$ ${\mu}$m.  
The parameters $\Lambda_{m}$ and $L_m$ from the linear theory are given by following formulae:
\begin{eqnarray}
\Lambda_{m}/v_z&=&\frac{2\pi}{\sqrt{m(m^2-1)}} \sqrt{\frac{\rho a^3}{T}}, \label{eq2}\\
L_{m}/v_z&=&\frac{1}{2m(m-1)}\frac{{a}^{2}}{\nu}, \label{eq3}
\end{eqnarray}
where $T, \rho$ and $\nu$ are the surface tension, the density and the kinematic viscosity of the liquid, respectively, and ${v}_{z}$ is the speed of the jet.
The quantities $\Lambda_m/v_z$ and $L_m/v_z$ are the oscillation period and the decay time, respectively.

Typically the jet is operated at ${v}_{z}\sim10-20$ m/s. 
For ethanol at 20$^{\circ}$C, $T=2.23\times{10}^{-2}$ N/m, $\rho=789$ kg/m$^3$, and $\nu=1.52\times{10}^{-6}$ m$^2$/s.
With these values we obtain the period of quadrupolar oscillation $\Lambda_2/v_z=28.0~\mu$s from Eq.\ (\ref{eq2}).
It agrees well with the observed value within the experimental error.
The decay time obtained from Eq.\ (\ref{eq3}) is $L_2/v_z=37.0~\mu$s.
Comparing the quadrupolar and octapolar modes, we obtain $L_2/L_4=6$ and $\Lambda_2/\Lambda_4=\sqrt{10}$.
During one period of the quadrupolar oscillation, $\Lambda_2=300-600 \mu$m, the quadrupolar amplitude is halved whereas the octapolar amplitude decays by $2$ orders of magnitude ($e^{-4.5}\simeq10^{-2}$). 
Therefore, after one quadrupolar period, the octapolar mode should become hardly noticeable according to the linear theory, 
and thus we expect that the cross-sectional shape of the jet at extreme positions D$n$ with $n>2$ [see Fig.~1(c)] would be approximated as
\begin{equation}
r/a\simeq 1+{\eta}(z)\cos2\theta,\label{eq4}
\end{equation}
where ${\eta}(z)$ is identified as the eccentricity $\epsilon$ of the cross-sectional shape, given by the relation
\begin{equation}
\epsilon=\frac{L_{\rm major}(z)-L_{\rm minor}(z)}{L_{\rm major}(z)+L_{\rm minor}(z)}.\label{eq5}
\end{equation}
where $L_{\rm major}(z)$ and $L_{\rm minor}(z)$ are the lengths of the major and minor axes at the observed extreme position, respectively.
Note one period of the lowest-order oscillation from the nozzle in Fig.\ 1(c) corresponds to extreme position D2.

\subsection{Inconsistency of linear analysis with spectroscopic observations}

The above linear analysis cannot account for the spectroscopic observations performed on our liquid jet. 
Before discussing the details of the inconsistency, let us first consider how those observations are made.

\begin{figure*}
\centering
\includegraphics[width=6in]{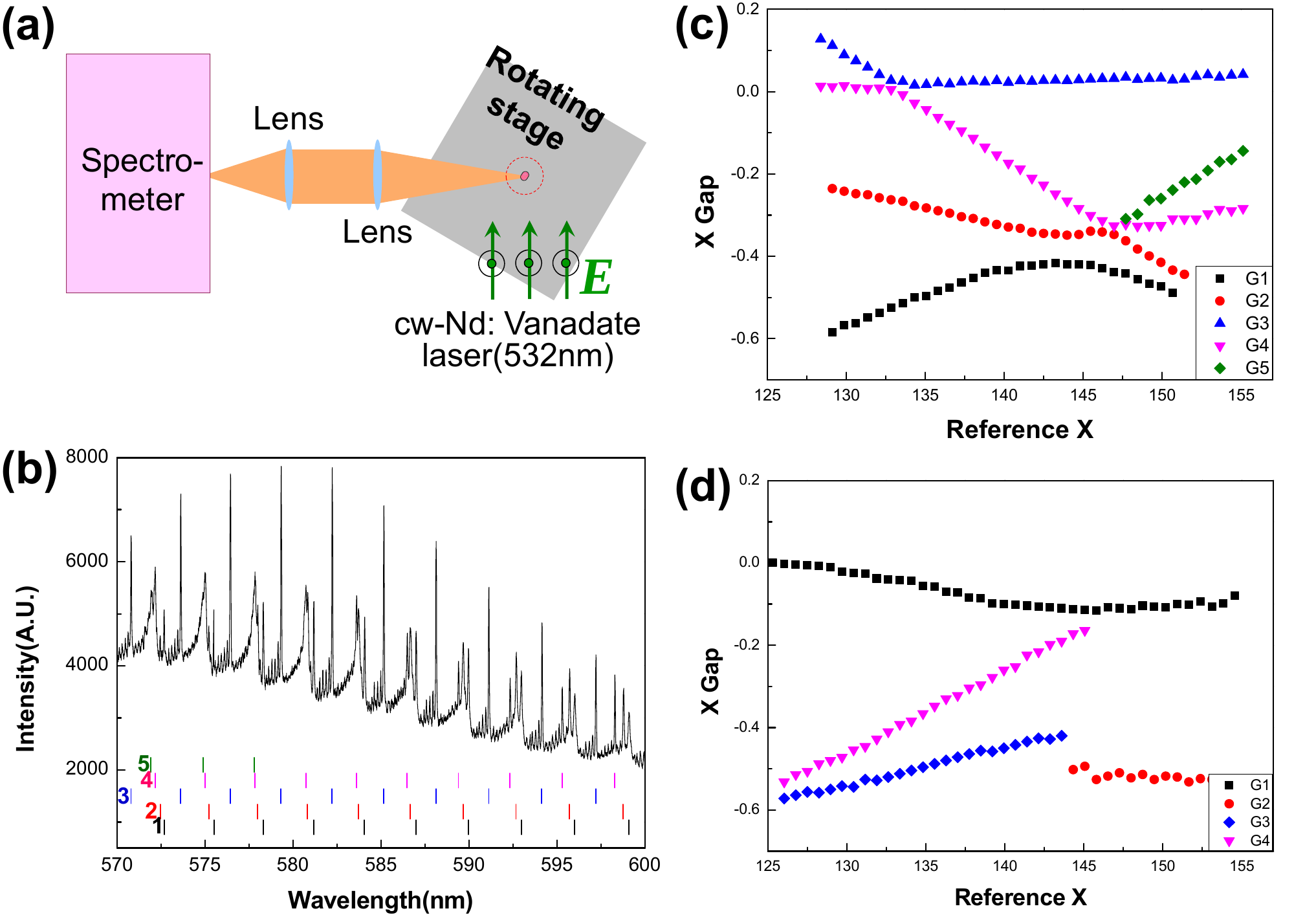}
\caption{
(a) Experimental setup for CMF spectroscopy (top view). 
The incident laser is polarized in the direction parallel to the jet column. 
(b) The observed CMF spectrum and identification of five groups (labeled as 1, 2, \ldots, 5) of cavity modes. 
The spectrum is measured at D4 with a jet-ejection pressure of 1.408 bar. 
The jet medium is ethanol doped with Rhodamine590 dye at a concentration of 0.05 mM/L. 
The mean radius of the jet is $a=14\pm1~{\mu}$m. 
Positions of the cavity modes are marked by vertical bars colored differently for each mode group.
(c) Experimentally observed mode evolution diagram when $\epsilon=0.19\pm0.02$. 
(d) Mode evolution diagram obtained by the wave calculation for the boundary profile given by Eq.~(\ref{eq4}) with $\eta=\epsilon=0.19$. 
Legends G1, G2, etc denote cavity mode groups.
}
\label{fig3}
\end{figure*}

A small $z$ segment of a few micron thickness of the liquid column at an extreme position acts as a two-dimensional dielectric cavity for the optical wave. 
The boundary of the cavity is given by the cross-sectional (perpendicular to the direction of the jet advance) shape of the jet. 
The liquid contains dye molecules. When the small $z$ segment of the jet column is excited by a pump laser as illustrated in Fig.~\ref{fig3}(a), the fluorescence from dye molecules is enhanced at the cavity resonances as shown in Fig.~\ref{fig3}(b). 
This enhancement is the cavity quantum electrodynamics effect \cite{Berman}.
In this cavity-modified fluorescence (CMF) spectrum from the microjet cavity, we typically observe $4-6$ groups of cavity resonances or modes. 
Each mode group is a sequence of resonances with a well-defined free spectral range (FSR).

The CMF spectrum in Fig.~\ref{fig3}(b) was measured with cavity eccentricity $\epsilon=0.19\pm0.02$, which was calculated by using Eq.~(\ref{eq5}) with the measured $L_{\rm major}$ and $L_{\rm minor}$ values from the microscope image.
The CMF spectrum was taken in a spectral range from $540$ nm to $660$ nm and all cavity modes were identified.
The wavelengths of those modes were then converted into a {\em normalized} size parameter $X=2{\pi}na/(n_0{\lambda})$, where $n$ is the wavelength- and temperature-dependent refractive index of the cavity medium (ethanol) and $n_0=1.361$, the value of $n$ at 610 nm and 20$^{\circ}$C. 
Their relative size parameters (labeled as X Gap) are then displayed in Fig.~\ref{fig3}(c) with respect to reference size parameters. 
The reference size parameters (displayed on the horizontal axis) are given by an arithmetic sequence with a mean FSR of all mode groups. 
The diagram obtained in this way as in Fig.~\ref{fig3}(c) is called the `mode evolution' diagram \cite{Lee09PRA}. 
It clearly shows intermode interactions, {\em i.e.}, crossing and avoided crossing of modes.

These observed modes are then compared with those by numerical wave calculation. The wavelength and linewidth of a cavity resonance are obtained by solving the Maxwell equations for a cylindrical dielectric medium whose 2-D boundary profile is given by Eq.~(\ref{eq4}) (quadrupole deformation only) with $\eta=\epsilon=0.19$.
We use the boundary element method for the wave calculation \cite{Wiersig}. 

For visual comparison, the mode evolution diagram obtained from the wave calculation is displayed in Fig.~\ref{fig3}(d). If Equation (\ref{eq4}) gives a correct description of the cross-sectional shape of the jet, the experimental and theoretical mode evolution diagrams should match each other. 
Instead, we find a big discrepancy. In the experiment we observe $5$ distinct high-$Q$ (quality factor) mode groups whereas we only find 4 such groups in the wave calculation. 
Moreover, the way the modes interact with each other shows a complete discrepancy: the relative spectral positions of all resonance-modes are quite different in both cases. 

One may suspect that the uncertainty in the eccentricity might be responsible for such a big discrepancy. 
We find, however, that in order to see 5 high-$Q$ mode groups in the wave calculation the eccentricity has to be less than $0.14$, which is beyond our experimental uncertainty. 
Moreover, even with such a low eccentricity the mode evolution diagram from the wave calculation look completely different from the experimental mode evolution diagram.
We then have to conclude that the cross-sectional profile of the jet is not given by Eq.~(\ref{eq4}).

What is then the actual cross-sectional shape of the jet?
The linear theory as an approximate description can explain the overall features of the jet such as axis switching and oscillation period. 
Therefore, it is natural for us to assume that the quadrupolar component is truly dominant in the actual boundary profile.  
There may exist a small shape perturbation which might be responsible for the different resonance mode interactions. 
One possible candidate for such a small perturbation is an octapolar component. 
The hexapolar component with $\cos 3\theta$ dependence is ruled out by the observed perpendicularity of the major and the minor axes or two-fold axial (D$_2$) symmetry of the observed cross-sectional shape.

One may then wonder how such a small perturbation can affect the cavity resonances so seriously. Considering {\em particle motion} in a deformed billiard is helpful to answer this question. In quantum chaos, it is a well-established fact that there exists a close connection between a classical trajectory of a particle (in our case, ray trajectory) and a wave-mechanical solution (cavity resonances) for a given billiard shape \cite{Noeckel97, Shim08}. 
It is demonstrated in Appendix A that 
a long-time particle or ray dynamics in a quadrupolar billiard is very sensitive to a shape perturbation. 
If experiments are performed with low-$Q$ cavities, which is often the case with etched semiconductor microcavities, long-time dynamics hardly survive to affect the cavity resonances much. 
On the other hand, if we deal with high-$Q$ resonances, such as in dynamical tunneling and mode evolution, long-time dynamics becomes important and thus a small error in the boundary profile can lead to a considerable difference \cite{Harald04, Tanaka07,  Lee09PRL, Shinohara, Yang10}.
By this reason, the spectroscopy of high-$Q$ resonance modes can serve a very sensitive measure of the boundary shape of the $2$-D optical billiard or the cross-sectional shape of our liquid jet.

\begin{figure}
\centering
\includegraphics[width=3.4in]{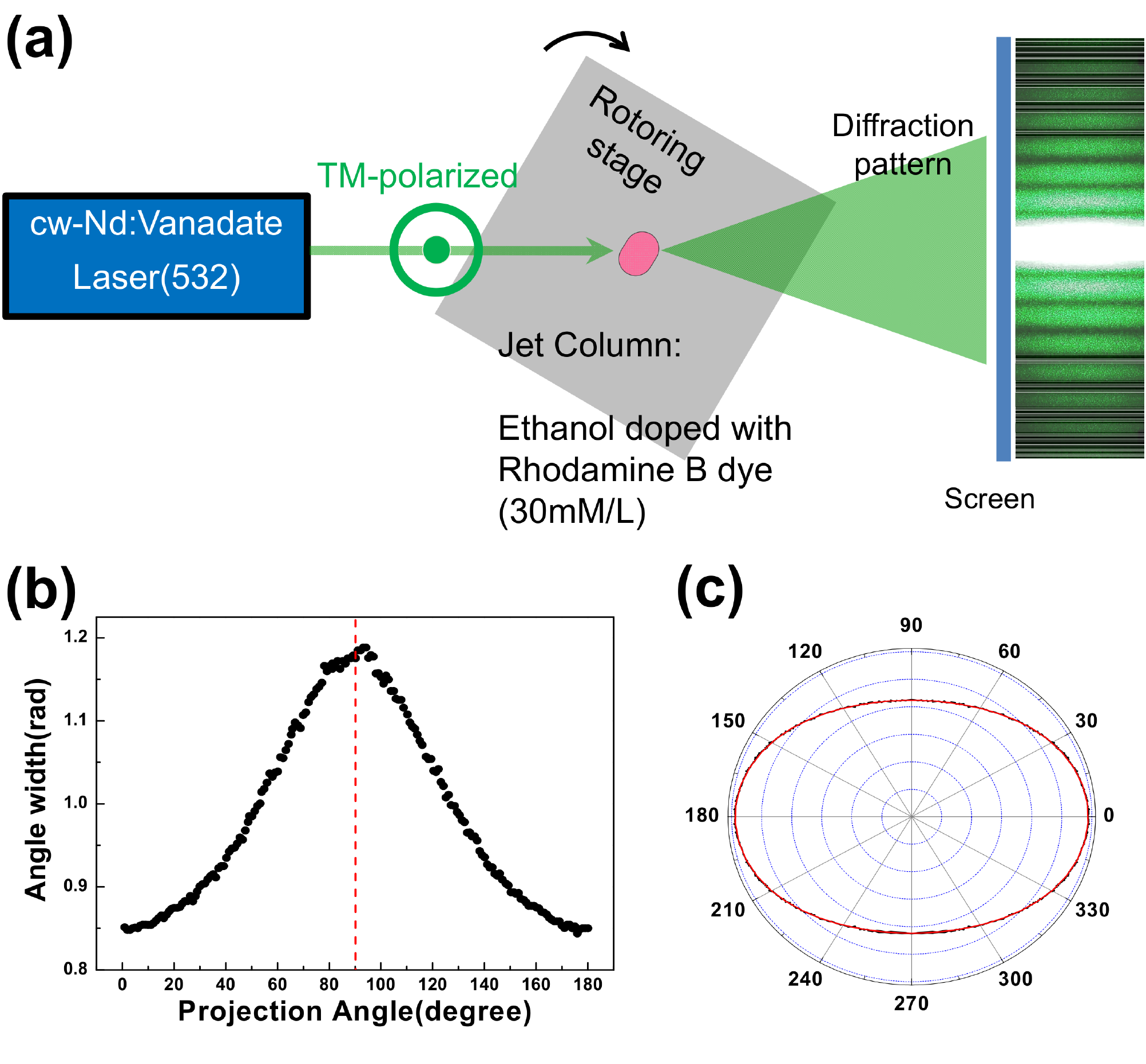}
\caption{(a) Top view of our experimental setup and a real image of the diffraction pattern. 
(b) An example of the angle width function $\Delta\Theta$. 
(c) Reconstructed profile from (b).}
\label{fig4}
\end{figure}

\section{Experiments and results} \label{sec4}

\subsection{Measurement of the boundary profile}


In our previous studies, we developed an optical technique to map out the cross-sectional profile of an opaque cylindrical object from its forward diffraction patterns with a resolution of $0.1\%$ of the mean diameter of the object \cite{Moon08}.
This technique is particularly useful for fluidic columns like a liquid microjet, where the conventional electronic imaging techniques such as scanning electron microscopy or scanning tunneling microscopy cannot be applied.
We will use this technique to determine the cross-sectional boundary profile of our liquid jet.
Below we recapitulate its working principle briefly.

Let us consider the forward diffraction pattern of an incident plane wave by an opaque cylindrical object.
It can be shown that the angular separation $\Delta\Theta$ between two adjacent local minima of the diffracted-wave intensity distribution near the forward direction is given by the following simple relation.
\begin{equation}
\Delta\Theta=\frac{\lambda}{2w}\label{eq7}
\end{equation}
where $2w$ is the projected full width of the object in the forward direction and $\lambda$ is the wavelength of the incident plane wave. 
We can then reconstruct the cross-sectional profile of the object by measuring $w$ at various projection angles, covering a range of at least $180^{\circ}$ for an object of two-fold axial symmetry.
Variation of the projection angle is conveniently done by rotating the cylindrical object on a rotatable stage for a fixed incident plane wave.
The details of the boundary profile reconstruction procedure is reported in Ref.~\cite{Moon08}.

In our experiment depicted in Fig.~\ref{fig4}(a), a liquid jet assembly is mounted on a rotary stage and the jet vertically ejected is illuminated by a cw laser beam at $532$ nm (or 514 nm in some cases), which is focused by a cylindrical lens with a focal length of $7$ cm down to a vertical waist of $3$ $\mu$m on the jet. 
The diffraction pattern of the laser beam by the jet column is then examined on a screen located at $2.4$ m from the jet.
The mean radius of the jet column is about $15$ $\mu$m while the horizontal waist of the laser beam is $2.3$ mm, fully covering the jet column horizontally.
About $5-6$ local minima lying within $10^{\circ}$ from the forward direction are used to obtain a mean value of $\Delta \Theta$ for a given rotation angle $\theta$ of the jet assembly.
The typical value of the resulting $\Delta\Theta$ is about 1 degree.

The requirement of opaqueness is satisfied by doping the liquid jet with dye molecules until the optical density due to absorption for a laser beam  going through the diameter exceeds $6.9$. 
For this, Rhodamine $590$ dye, whose absorption cross section $\sigma$ at $\lambda$=$532$ nm is $3 \times 10^{-21}$ m$^2$, is dissolved in ethanol at a molar density of $23$ mM/L.
The boundary profile measurement is repeated for various initial conditions such as different ejection pressure, types of solvent and mean radii at several critical positions D$n$ ($n>2$) on the jet. 

All boundary profiles we reconstructed exhibit $2$-fold axial symmetry. 
The profiles are dominated by a quadrupolar component. 
The second largest component is a octapole component.
The magnitude of the octapole component in almost all cases cannot be explained by the linear theory.
For example, $\Delta\Theta$ as a function of $\theta$ in Fig.~\ref{fig4}(b) and its reconstruction profile in Fig.~\ref{fig4}(c) are obtained at an ejection pressure of $2.6$ bar with $r=15~{\mu}$m and $v_z=16$ m/s at D$3$, which is located at 860 $\mu$m (corresponding to about $2\Lambda_2$) from the orifice.
The relative magnitudes of the quadrupole and octapole components there are $23.9 ({\pm} 0.3)$\% and $2.6 (\pm 0.2)$\%, respectively.
The magnitudes of the higher poles are smaller than $0.1$\%, which is within the error range of the present technique.
If the linear theory were correct, the observed magnitude of the octapole component should come from an initial relative magnitude exceeding $15000$\%, which is simply impossible.

\begin{figure}
\centering
\includegraphics[width=3.4in]{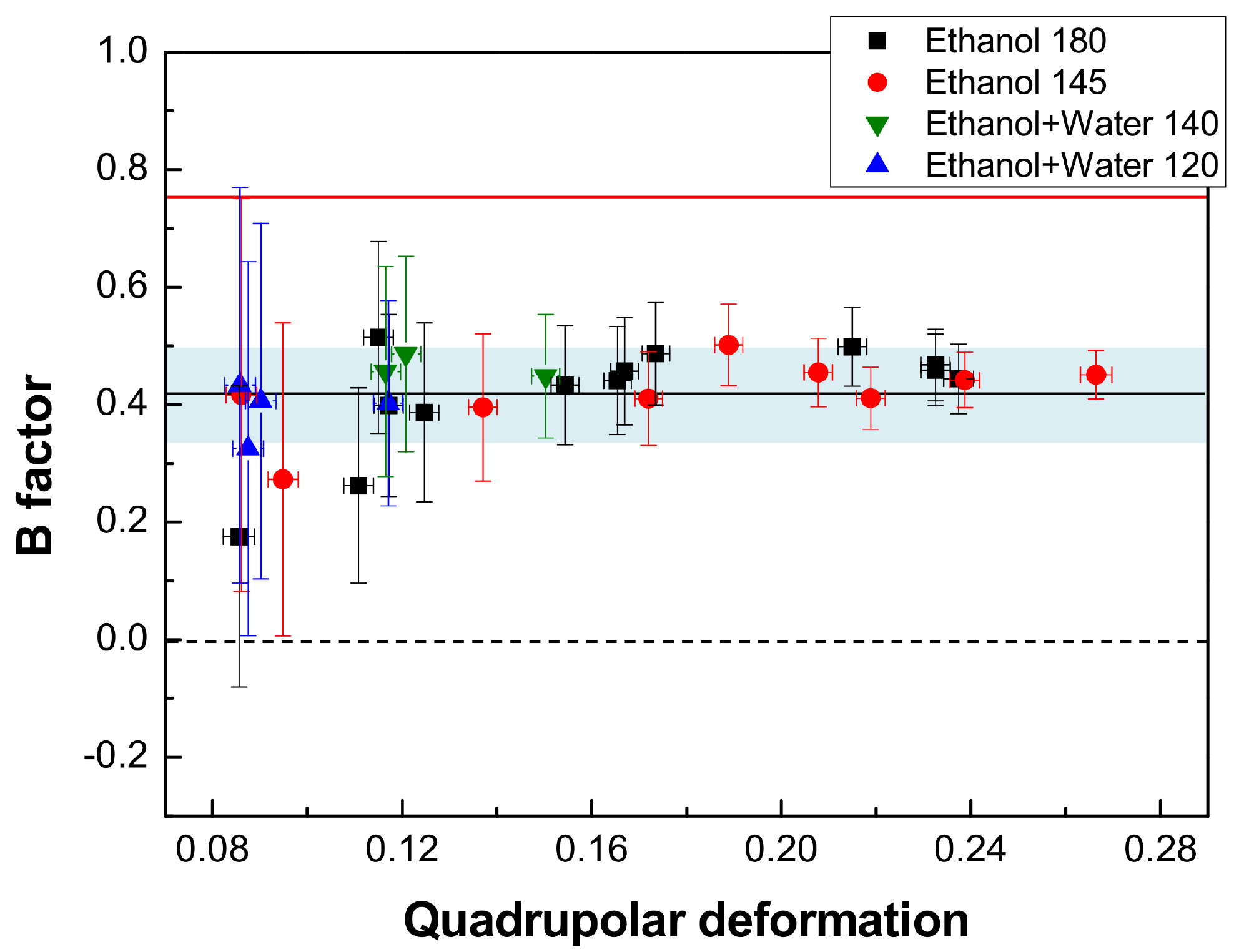}
\caption{
$B$ values for various deformations and materials. 
The black-filled  squares and red-filled circles are data of ethanol, while the blue- and green-filled triangles are that of ethanol-water $50$:$50$ mixture. 
Numbers are the normalized size parameter $X$. 
The averaged $B$ factor (black line) is clearly distinguished from the $B$ factor for an ellipse (red line at $B=0.75$) and that for a quadrupole (dotted line at $B=0$). 
The fitting error is indicated as a shaded region.
}
\label{fig5}
\end{figure}

\begin{figure*}
\centering
\includegraphics[width=6in]{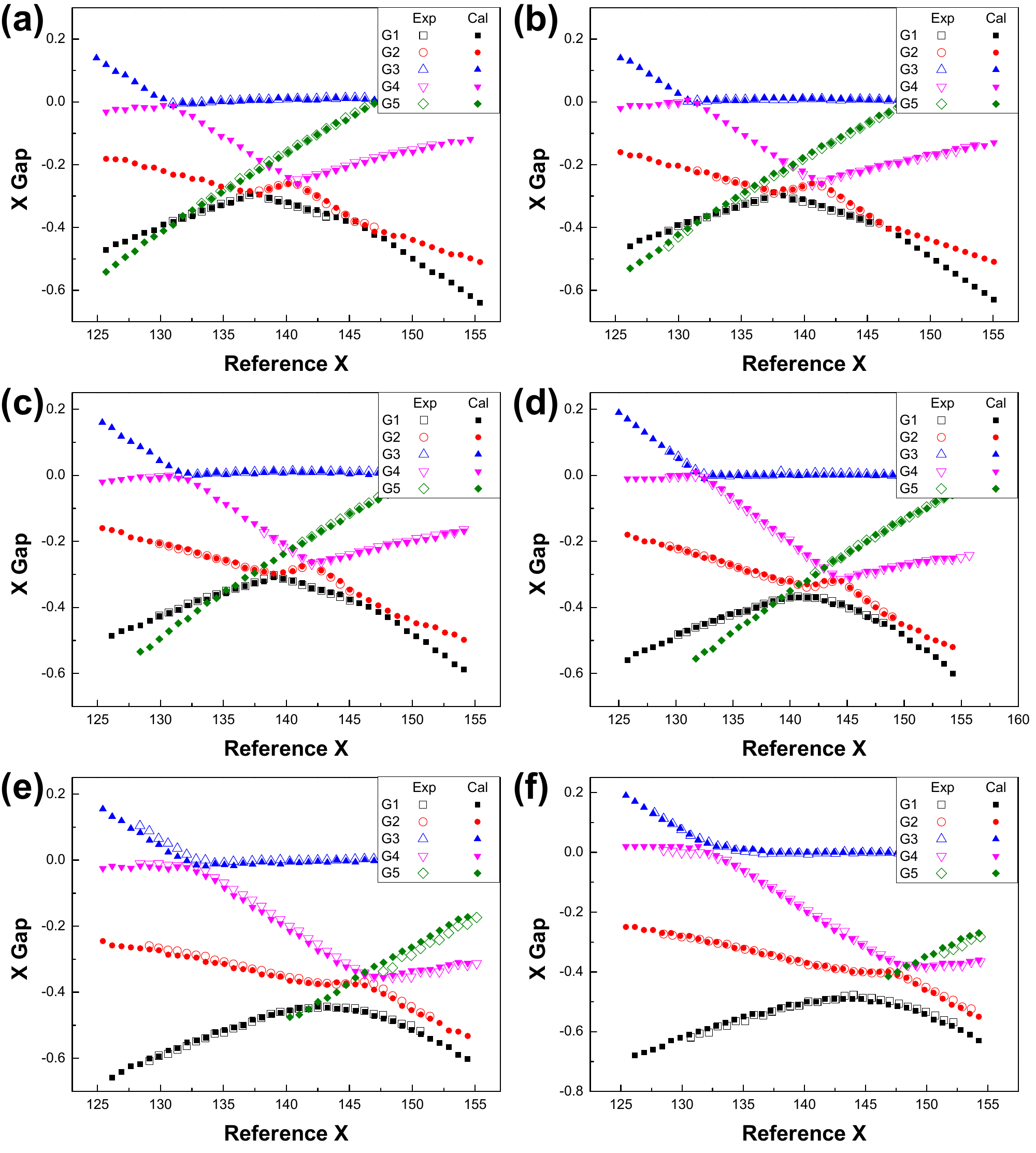}
\caption{
Matching the mode evolution diagrams between experiment (denoted by `Exp') and wave calculations (denoted by `Cal'). 
Experimental conditions are expressed as `(jet pressure in bar, extreme position D$n$, jet temperature in $^{\circ}$C)'. Jet mean radius $a$ for each case is determined as a fitting parameter.
(a) Experiment  at (1.250, D5, 16.9) and the wave calculation with ${\eta}_{1}=0.095$. $a$=13.79$\pm$0.01 ${\mu}$m.
(b) Experiment  at (1.402, D5, 16.9) and the wave calculation with ${\eta}_{1}=0.105$. $a$=13.71$\pm$0.01 ${\mu}$m.
(c) Experiment  at (1.326, D5, 21.4) and the wave calculation with ${\eta}_{1}=0.130$. $a$=13.64$\pm$0.01 ${\mu}$m.
(d) Experiment  at (1.188, D4, 20.3) and the wave calculation with ${\eta}_{1}=0.160$. $a$=13.61$\pm$0.01 ${\mu}$m. 
(e) Experiment  at (1.408, D4, 19.5) and the wave calculation with ${\eta}_{1}=0.185$. $a$=13.55$\pm$0.01 ${\mu}$m.
(f) Experiment  at (1.536, D4, 19.1) and the wave calculation with ${\eta}_{1}=0.200$. $a$=13.50$\pm$0.01 ${\mu}$m. The data in (e) is the same as the data in Fig.~3(c).
}
\label{fig6}
\end{figure*}

Another noticeable point is that there exists a relation between the magnitude $\eta_1$ of the quadrupolar component and the magnitude $\eta_2$ of the octapolar component as summarized by
\begin{equation}
\eta_{2}\simeq{B}{{\eta}_{1}}^{2}\label{eq9},
\end{equation}
where $B$ is a certain constant to be called `$B$ factor' below.
For an ellipse, from Eq.~(\ref{eq6}), we have $B=3/4$.
The observed boundary profile of the liquid jet also shows a similar relation, but with a different $B$ factor.
Figure \ref{fig5} summarizes the $B$ factors extracted from the observed boundary profiles under different experimental conditions such as mean radii, types of solvent and eccentricity.
Interestingly, they are centered around a common value $B=0.42\pm0.08$, if we require $B$ to be a constant, despite the different experimental conditions.
The boundary profile can then be described by the following empirical equation:
\begin{equation}
r/a\simeq 1+{{\eta}_{1}}\cos2\theta+0.42{{\eta}_{1}}^{2}\cos4\theta. \label{eq10}
\end{equation}

\subsection{Spectroscopic confirmation of  the $B$ factor}

In the preceding section, the spectroscopic observation was presented as a counter evidence against the linear analysis. 
Now, for the boundary profile given by the empirical formula Eq.~(\ref{eq10}) with ${\eta}_{1}=0.185\pm0.003$, the experimental mode evolution diagram in Fig.~\ref{fig3}(c) is in good agreement with the wave calculation result.
This $\eta_1$ value corresponds to an eccentricity $\epsilon$ of $0.182\pm0.003$, which is consistent with the measured eccentricity of $0.19\pm0.01$ for our liquid jet within the experimental error.
Besides, in Fig.~\ref{fig6} five more mode diagrams are presented for a wide range of ${\eta}_{1}$.

\begin{figure*}
\centering
\includegraphics[width=6in]{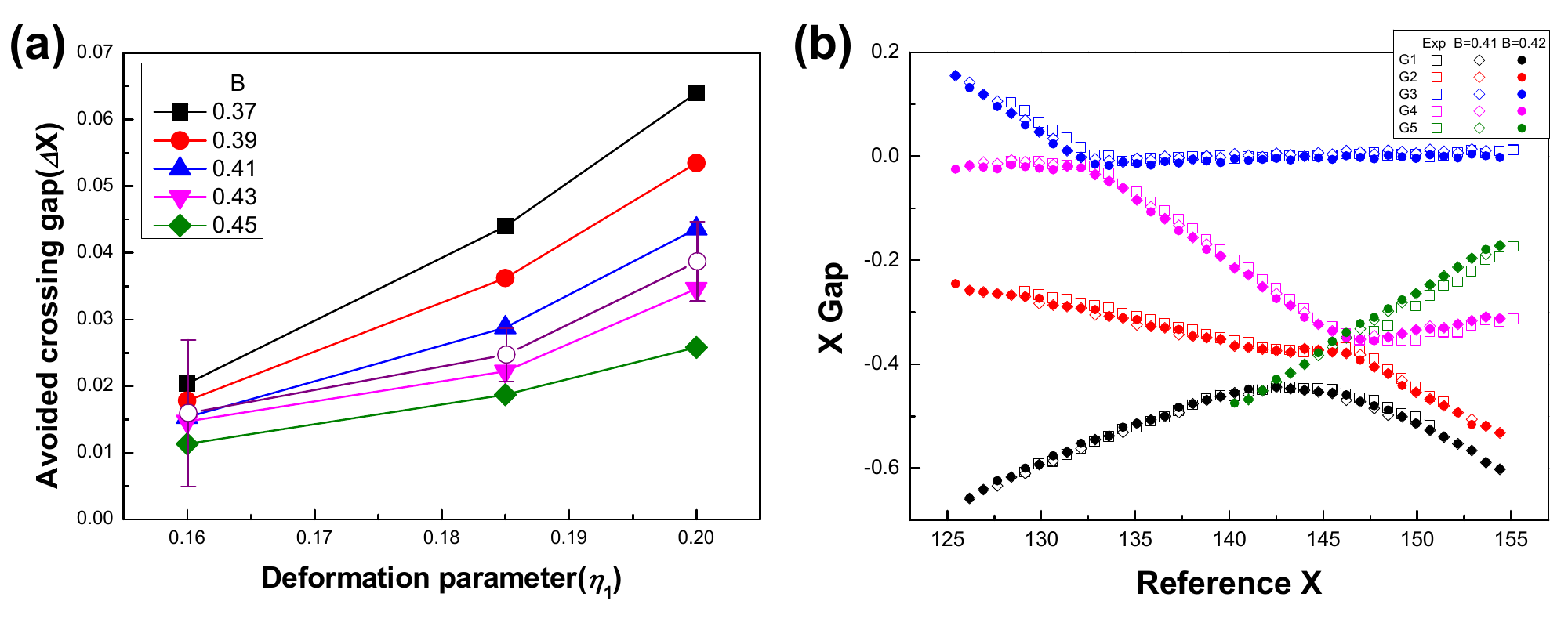}
\caption{
(a) $G2-G4$ avoided crossing gaps from the experiment (open circles with error bars) is compared with those from the wave calculation with the $B$ factor varied from $0.37$ to $0.45$ forn $\eta_1$=0.160, 0.185, 0.200. 
(b) Mode evolution diagrams for $B=0.41$ and $B=0.42$ for ${\eta}_1=0.19$.
}
\label{fig7}
\end{figure*}

One can evaluate the tolerance in the $B$ factor value of 0.42 by examining the difference between the observed mode evolution diagram and the calculated one as a function of $B$ around the empirical value of 0.42. 
For quantitative analysis, we pay attention to the avoided crossing gap between mode groups 2 and 4, which are denoted by G2 and G4, respectively, in Figs.~\ref{fig6} and \ref{fig7}.
The G2-G4 avoided crossing gap for $\eta_1=0.185$ is calculated from the wave equation for various $B$ factor values from $0.37$ to $0.45$ in Fig.~\ref{fig7}(a). 
We find that the experimental avoided crossing gap is well fit by the $B$ factor value of $0.414\pm0.011$. 
This is also supported in Fig.~\ref{fig7}(b), where both wave calculations with B=0.41 and 0.42 well reproduce the experimental mode evolution diagram for $\eta_1$=0.185.


We can now regard the empirical formula Eq.~(\ref{eq10}) as a correct description of the actual cross-sectional boundary profile of our liquid jet. 
The remaining question is then how we can derive the empirical formula from the hydrodynamics theory \cite{Tsamopoulos83} by going beyond the linear approximation.

\section{Nonlinear analysis} \label{sec5}

\subsection{Bohr's analysis}

In $1909$, Niels Bohr at the age of 23, before inventing the hydrogen model, considered the liquid jet problem including nonlinearity \cite{Bohr}.
Conceiving the limitation of the linear approximation in the previous study by P.\ O.\ Pederson \cite{Pederson}, he derived an expression for the surface oscillation of  a non-viscous liquid jet with a few lowest-order components while taking the nonlinear interactions among them into account \cite{note2}. 

When we rewrite his result in terms of our notations, we obtain
\begin{eqnarray}
r/{{a}_{0}}&=&1-\frac{1}{8}{\eta}^{2}-\frac{1}{8}{\eta}^{2}\cos2kz+{\eta}\cos2\theta\cos{kz}\nonumber\\
& &+\frac{1}{4}{\eta}^{2}\cos4\theta+\frac{1}{6}{\eta}^{2}\cos4\theta\cos2kz, 
\label{eq11}
\end{eqnarray}
where $k$ is the wave vector in the propagation direction, and ${{a}_{0}}$ is the mean radius when $\eta=0$.
Considering only nodal ($\cos{kz}=\cos2kz=1$) and anti-nodal ($\cos{kz}=-1, \cos2kz=1$) positions, and taking the major axis of the shape as the $\theta$ axis in the $r$-$\theta$ coordinates, Eq.~(\ref{eq11}) is simplified as
\begin{equation}
r/{{a}_{0}}=1-\frac{1}{4}{\eta}^{2}+\eta\cos2\theta+\frac{5}{12}{\eta}^{2}\cos4\theta. \label{eq12}
\end{equation}
From the flux conservation of the jet with its negligible speed change, we find 
\begin{equation}
a/a_0\simeq 1-\frac{{\eta}^{2}}{4}.
\label{eq_AreaConservation}
\end{equation}
Including this effect, the empirical formula Eq.~(\ref{eq10}) can be written as
\begin{equation}
r/{{a}_{0}}\simeq 1-\frac{1}{4}{\eta}^{2}+\eta\cos2\theta+0.42{\eta}^{2}\cos4\theta, \label{eq10'}
\end{equation}
which is almost the same as Bohr's result.
Only difference is the numerical factor 0.42 compared to a rational number $5/12=0.41\dot{6}$.
The calculation of Bohr's gives the theoretical $B$ factor of 5/12, which agrees well with our empirically $B$ factor within the experimental error.

\subsection{Limitation of Bohr's analysis and our extension} \label{correction}

In Bohr's analysis, the $z$-dependence is treated quite simply: the surface motion is regarded as a two-dimensional oscillation and its time dependence is simply replaced by the z-dependence with an assumption that the fluid velocity is constant.
This treatment is by no means rigorous by the following reasons.
First, under gravity, the fluid velocity is not constant. Particularly, if the jet is ejected vertically upward like in our experimental setup, the velocity slows down.
Second, the conservation of the flux dictates the velocity has to change periodically. The sectional area given by Eq.~(\ref{eq12}) is $\pi [1+\frac{1}{4}{\eta}^{2}(1-\cos2kz)]$, having $z$ dependence. 
Therefore, the velocity also has to change as $z$ in order keep the flux conserved.

If we limit ourselves to the region where the surface oscillations have not decayed sufficiently, {\em i.e.}, at several oscillation wavelengths away from the orifice, the change of velocity due to gravity is estimated to be negligibly small for our jet (less than $10^{-4}$ of the initial velocity). 
The velocity change due to the flux conservation, however, amounts to $1\%$ of the initial velocity, in the same order of magnitude as the octapole component in the typical boundary profile.
This is not small enough to be neglected, and it may affect the boundary profile.
Consequently, we cannot rule out the possibility that the agreement between the experimental result Eq.~(\ref{eq10}) or Eq.~(\ref{eq10'}) and Bohr's analysis Eq.~(\ref{eq12}) might be accidental under the particular conditions that the experiment was done.

In order to resolve this problem, we have extended Bohr's inviscid 2-D analysis into 3-D by including the effects of gravity and flux conservation. 
We have then obtained the coefficient attached to $\eta^2 \cos 4\theta$ as
\begin{equation}
B=\frac{5}{12}-\frac{232}{3240}{{a}_{0}}^{2}k^{2}\simeq 0.417-0.07{{a}_{0}}^{2}k^{2}. \label{eq34}
\end{equation}
The detailed derivation is  presented in Appendix B. The term containing $({{a}_{0}}k)^2$ is very small in our experiment:
$0.07{{a}_{0}}^{2}k^{2}\simeq0.42T/{\rho{{a}_{0}}{v_{z}}^{2}}\simeq 0.8$ [m$^{2}$/s$^{2}]/({v_{z}}^{2})\sim 10^{-3}$ considering that the velocity ${v}_{z}$ is typically $10-20$ m/s.
Experimentally, such a small correction is obscured by our experimental error. 
Therefore, our calculation confirms that the nonlinear analysis by N.~Bohr is adequate for describing the motion of our liquid microjet. However, if $a_0k$ is comparable to or larger than unity such as in the case of garden hose or home faucet, the adequateness of Bohr's analysis is not guaranteed and we expect that our extension should be applied.

\section{Conclusion} \label{sec6}

We have measured the cross-sectional shapes of a surface oscillating liquid jet at its extreme positions by using the optical forward diffraction method and determined the magnitude of their octapole components. 
The normalized magnitude of the octapole component was found to be empirically given by $B\eta^2$ with $\eta$ the magnitude of a quadrupole component and $B=(0.42\pm0.08)$.
We also measured the resonance mode spectra of a two-dimensional cavity formed by a cross-sectional segment of the liquid jet at various extreme positions.
The observed spectra agreed well with wave calculation assuming a proportionality coefficient $B=0.415\pm0.010$.
Combining these two results we were able to confirm the long-standing prediction of Niels Bohr's in his nonlinear theory of fluidic surface oscillation for the first time. 

Our finding can be utilized in precise measurement of surface tension especially when the material is in a non-equilibrium state such as being sprayed, spun, or blown.
Moreover, since our results provide an accurate shape formula for the surface oscillation of a fluidic jet,
one can match experimental data such as mode spectrum and output directionality from a given fluidic microjet cavity more accurately with theoretical ray and wave simulations. 
One can also predict mode dynamics for arbitrary cavity deformation in a fluidic microjet cavity in advance, and thus can help designing various photonics experiments based on intermode interactions.
One example is experimental verification of an enhanced gain due to the diverging Petermann factor near an exceptional point arising from intermode interactions. 
Fluidic microjet cavities can be incorporated into optofluidic systems \cite{Fainman} as optical resonators whose mode characteristics can be accurately designed with our results. Such systems can be a versatile platform for biosensors \cite{Armani}, optical filters and mode switchers \cite{Ilchenko}.

This work was supported by the Korea Research Foundation (Grant No.\ 2013R1A1A2010821).

\appendix

\section{Effect of a small perturbation in RAY dynamics} \label{appedix-A}

It is known that non-regular boundary shapes such as quadrupole can induce chaotic dynamics in the classical particle description and also result in the interactions among resonance modes in the wave description. 
In addition, it is known that even quite a small perturbation can change the particle dynamics significantly in a partially chaotic system. 
Considering the close connection between classical dynamics and wave solutions, we expect a small perturbation of a right type may explain the above discrepancy with the spectroscopic observations.

\begin{figure}
\centering
\includegraphics[width=3.4in]{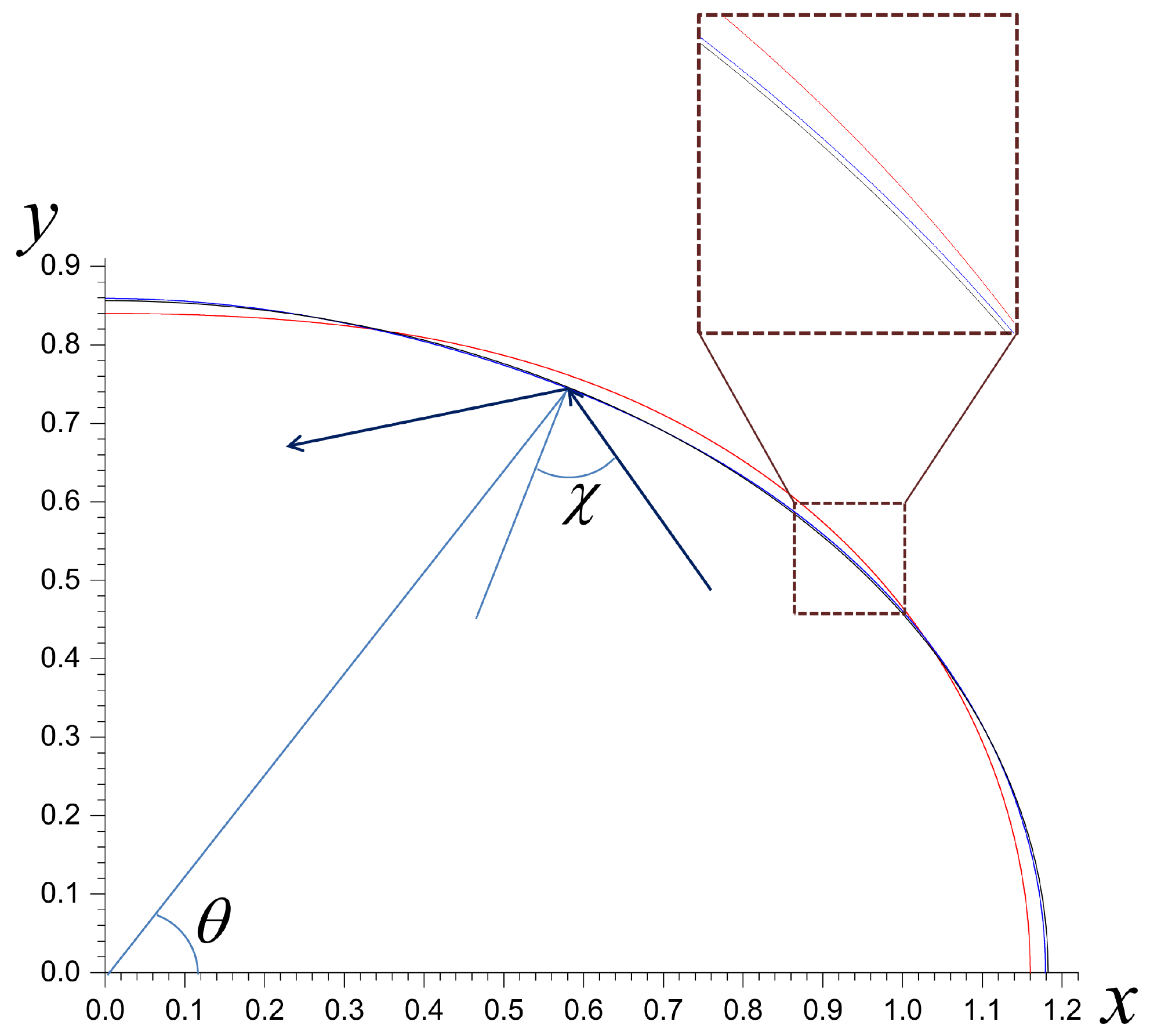}
\caption{
Comparison of an elliptical billiard (black curve), a semi-elliptical billiard (blue curve) and a quadrupolar billiard (red curve) with the same eccentricity $\epsilon=0.16$ and the same enclosed area. 
When a particle is reflected off the boundary, its angular position is denoted by $\theta$ with an reflection (or incident) angle $\chi$. 
}
\label{fig8}
\end{figure}

A regular billiard which has a profile similar to a quadrupole is an elliptical billiard, whose boundary can be expressed as an infinite series of multipolar components as
\begin{equation}
r/a \simeq 1+\eta\cos2\theta+\frac{3}{4}{\eta}^{2}\cos4\theta+\frac{5}{8}{\eta}^{3}\cos6\theta\cdots \label{eq6}
\end{equation}
The shape difference between a pure quadrupole and the ellipse with the same eccentricity is quite small, amounting to at most $1/100$ of the mean diameter $2a$ in the deformation range of $0<{\eta}<0.2$ as seen in Fig.\ \ref{fig8}.
Beyond $\eta=0.2$, the convexity of quadrupole is broken like peanut due to the contraction along the minor axis. 
However, internal ray dynamics in the elliptical billiard are quite different from those of the pure quadrupolar billiard.
It is because a small initial difference in particle trajectories can grow exponentially later on in the presence of chaos.

\begin{figure*}
\centering
\includegraphics[width=6in]{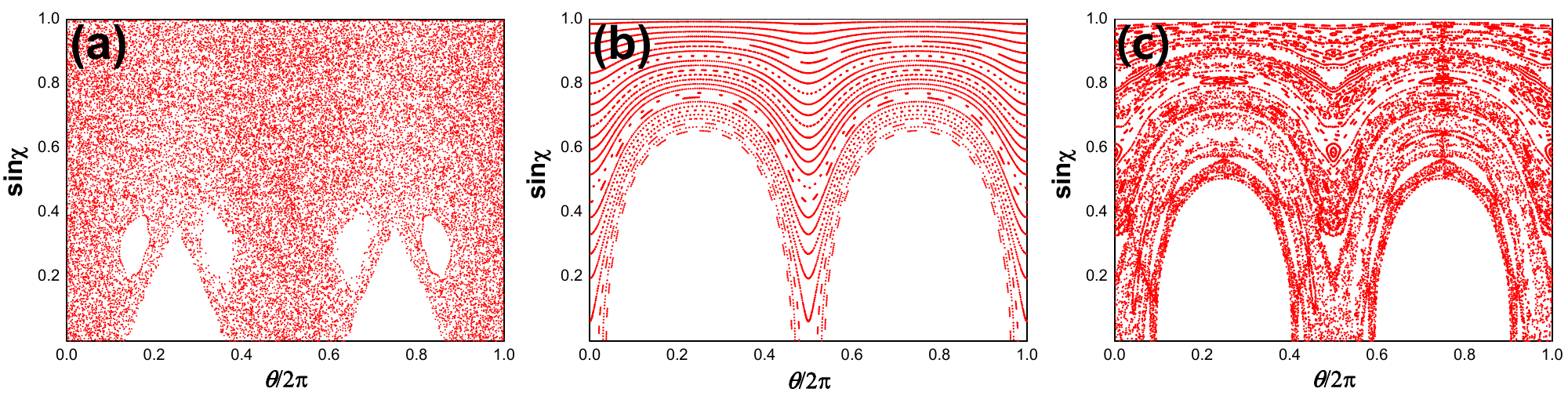}
\caption{
Poincare surface of sections for various billiards. 
(a) A pure quadrupole given by $1+0.16\cos2\theta$, 
(b) an ellipse and 
(c) a semi-ellipse, both with an eccentricity $\epsilon=0.16$.
}
\label{fig9}
\end{figure*}

Classical particle dynamics is conveniently analyzed in Poincar\'{e} surface of sections (PSOS), which is commonly used to visualize the particle motion in a $2$-dimensional billiard. 
PSOS is a phase-space diagram which is drawn iteratively in terms of reflection position ($\theta$) and reflection angle ($\chi$) of a particle upon bouncing off the boundary. 
Regular or quasi-periodic orbits are represented by concentration of dots around localized substructures such as islands or curves. 
Chaotic motion has no such concentration of dots, rather randomly but uniformly filling some regions called chaotic sea in the phase space.

Figure \ref{fig9}(a) shows the PSOS of a pure quadrupole while Fig.~\ref{fig9}(b) shows that of an ellipse, both drawn in terms of $\theta$ and $\sin\chi$ for $\epsilon=0.16$.
Almost the whole PSOS in Fig.~\ref{fig3}(a) is chaotic while the PSOS in Fig.~\ref{fig9}(b) consists of many periodic and quasi periodic orbits without chaos. 
Since it is practically impossible to fabricate a billiard with a perfectly elliptical or perfectly quadrupolar boundary, 
it is also meaningful to consider the PSOS of a truncated semi-ellipse with the same eccentricity for comparison as in Fig.~\ref{fig9}(c).
The profile of a semi-elliptical billiard is obtained by truncating the expansion in Eq.~(\ref{eq6}) beyond the second order of $\eta$. 
Though chaotic sea is present in considerable regions of the phase space, we still have curve-like or island-like structures corresponding to regular trajectories. 
From these considerations, we find that even quite a small difference in the boundary profile can change the particle dynamics significantly. 
In particular, degradation of the regular dynamics is striking when an ellipse is replaced with a semi-ellipse with the same eccentricity.

\section{Improved nonlinear analysis} \label{appendix-B}

In order to derive the expression for the $B$ factor more rigorously,
we have modified the nonlinear calculation of Niels Bohr to include the $z$-dependence up to the second order of $\eta$.
We include the viscous decay in terms of an exponential factor multiplied to an inviscid surface oscillation mode.
Then the surface motion can be treated by introducing the potential $\phi$ for the velocity $\mathbf{v}$, satisfying $\nabla\phi=\mathbf{v}$.
Master equations of motion derived from the Bernoulli's principle under the stationary condition are then written as follows in the cylindrical coordinates $(r,\theta,z)$:
\begin{equation}
{\nabla}^2\phi=0, \label{eq13}
\end{equation}
\begin{equation}
\frac{T}{{\rho}a}\left(\frac{a}{{R}_{1}}+\frac{a}{{R}_{2}}\right)+\frac{1}{2}{\left(\frac{{\partial}{\phi}}{{\partial}{r}}\right)}^{2}+\frac{1}{2{r}^{2}}{\left(\frac{{\partial}{\phi}}{{\partial}{\theta}}\right)}^{2}+\frac{1}{2}{\left(\frac{{\partial}{\phi}}{{\partial}{z}}\right)}^{2}=C, \label{eq14}
\end{equation}
\begin{equation}
\frac{{\partial}{\phi}}{{\partial}{r}}-\frac{1}{{r}^{2}}\frac{{\partial}{\phi}}{{\partial}{\theta}}\frac{{\partial}{r}}{{\partial}{\theta}}-\frac{{\partial}{\phi}}{{\partial}{z}}\frac{{\partial}{r}}{{\partial}{z}}=0, \label{eq15}
\end{equation}
where
$C$ is a constant and ${R}_{1}$ and ${R}_{2}$ represent two principal radii of curvature of the jet surface.
Equations (\ref{eq13})-(\ref{eq15}) account for the irrationality of the fluid, the Bernoulli's principle and the condition for unbroken surface, respectively.
The sum of the reciprocals 
of the principal radii is just the trace of the shape operator in differential geometry: 

\begin{eqnarray}
\frac{a_0}{R_1}+\frac{a_0}{R_2}&=&\left[-{r}^{3}\frac{{\partial}^{2}r}{\partial{z}^{2}}+2r\frac{{\partial}{r}}{{\partial}{\theta}}\frac{{\partial}{r}}{{\partial}{z}}\frac{{\partial}^{2}{r}}{{\partial}{\theta}{\partial}z} \right. \nonumber\\
& &-r\frac{{\partial}^{2}r}{{\partial}{\theta}^{2}}{\left(\frac{{\partial}r}{{\partial}z}\right)}^{2} 
-r{\left(\frac{{\partial}r}{{\partial}\theta}\right)}^{2}\frac{{\partial}^{2}r}{\partial{z}^{2}}  \nonumber\\
& & \left. -{r}^{2}{\left(\frac{{\partial}r}{{\partial}z}\right)}^{2} +{r}^{2}
+2{\left(\frac{{\partial}r}{{\partial}\theta}\right)}^{2} 
-r\frac{{\partial}^{2}r}{\partial{\theta}^{2}}\right] \nonumber\\
& & \left[{r}^{2}+{\left(\frac{{\partial}r}{{\partial}\theta}\right)}^{2}+{r}^{2}{\left(\frac{{\partial}r}{{\partial}z}\right)}^{2}\right]^{-\frac{3}{2}}.\label{eq16}
\end{eqnarray}
The surface motion is decomposed by the orders of magnitude in $\eta$:
\begin{eqnarray}
&&r/{{a}_{0}}={F}^{(0)}+{\eta}{F}^{(1)}+\frac{1}{2}{\eta}^{2}{F}^{(2)}, \label{eq17}\\
&&\phi={\phi}^{(0)}+\eta{\phi}^{(1)}+\frac{1}{2}{\eta}^{2}{\phi}^{(2)}. \label{eq18}
\end{eqnarray}
Then Eq.~\ref{eq16} becomes
\begin{eqnarray}
\frac{a_0}{R_1}+\frac{a_0}{R_2}&=&1-\eta \left[F^{(1)}+\frac{\partial^2 F^{(1)}}{\partial \theta^2}+a_0^2 \frac{\partial^2 F^{(1)}}{\partial z^2} \right] \nonumber\\
& &-\frac{1}{2}\eta^2 \left[ F^{(2)}+\frac{\partial^2 F^{(2)}}{\partial \theta^2} -2\left(F^{(1)}\right)^2 \right.\nonumber\\
& &-4F^{(1)}\frac{\partial^2 F^{(1)}}{\partial \theta^2} -\left(\frac{\partial F^{(1)}}{\partial \theta}\right)^{2}
+a_0^2 \left(\frac{F^{(1)}}{\partial z}\right)^2 \nonumber\\
& & \left. +a_0^2 \frac{\partial^2 F^{(2)}}{\partial z^2}\right].
\label{eq19}
\end{eqnarray}
The zeroth motion corresponds to an upward cylindrical flow of a circular cross section. 
The gravitation causes the vertical velocity to decrease as the height: ${v}^{2}={v_z}^{2}-2gz$, where $g$ is the gravitational acceleration and $v_{z}$ is the initial velocity.
At the same time the radius has to increase to conserve the flux. 
Therefore, $F^{(0)}$ and $\phi^{(0)}$ are given by
\begin{equation}
{F}^{(0)}=1+\frac{gz}{2{{v}_{z}}^{2}}, \label{eq20}
\end{equation}
\begin{equation}
{\phi}^{(0)}=\frac{gr^{2}}{4v_{z}}+{v}_{z}z-\frac{g}{2v_{z}}z^{2}. \label{eq21}
\end{equation}
The $r$-dependence of the velocity potential is required to satisfy the Laplace equation.
With these expressions substituted in Eq.~(\ref{eq14}), we can identify $C=T/(\rho {{a}_{0}})+v_{z}^{2}/2-gz$.
However, the terms containing the gravitational acceleration $g$ are too small to make noticeable contribution.
They will be thus neglected from now on.

The motion corresponding to the first-order terms is quite similar to the motion from the linear analysis. The master equations are:
\begin{equation}
{\nabla}^2{\phi}^{(1)}=0, \label{eq22}
\end{equation}
\begin{equation}
-\frac{T}{{\rho}{{a}_{0}}}\left[{F}^{(1)}+\frac{{\partial}^{2}{F}^{(1)}}{\partial{\theta}^{2}}+{{a}_{0}}^{2}\frac{{\partial}^{2} {F}^{(1)}}{\partial{z}^{2}}\right]+v_{z}\frac{\partial{\phi}^{(1)}}{\partial{z}}=0, \label{eq23}
\end{equation}
\begin{equation}
\frac{\partial{\phi}^{(1)}}{\partial{r}}-{{a}_{0}}v_{z}\frac{\partial{F}^{(1)}}{\partial{z}}=0. \label{eq24}
\end{equation}
A solution of the Laplace equation in the cylindrical coordinates can be written as a linear sum of Bessel's functions $J_{n}$.
Therefore above equations are solved by substituting
\begin{eqnarray}
{F}^{(1)}&=&\cos{n}\theta\cos{kz} \nonumber\\
{\phi}^{(1)}&=&M_{n}J_{n}(ikr)\cos{n}\theta\sin{kz}
\label{eq24-2}
\end{eqnarray}
where $n=1,2,3,\ldots$ and $M_n$ is a coefficient to be determined.
Equations (\ref{eq23}) and (\ref{eq24}) then yield
\begin{equation}
M_{n}=-\frac{T}{\rho{a}kv_{z}J_{n}(ik{{a}_{0}})}(n^{2}+k^{2}{{a}_{0}}^{2}-1)=\frac{i{{a}_{0}}v_{z}}{{J'}_{n}(ik{{a}_{0}})}.\label{eq25}
\end{equation}
The factor $(k{{a}_{0}})$ in the Bessel's function is noticeable. 
It is interpreted as an index number representing the ratio of the mean radius to the $z$-axial wavelength.
A circular cylinder is the case when $k{{a}_{0}}\rightarrow 0$.

Equation (\ref{eq25}) is further simplified when $k{{a}_{0}}\ll1$.
Using the approximated expression for the Bessel's function, the above relation can be approximated as
\begin{equation}
(k{{a}_{0}})^{2}=\frac{T}{\rho{{a}_{0}}^{3}v_z^2}n(n^{2}-1)\left[1+\frac{3n-1}{2n(n^{2}-1)}(k{{a}_{0}})^{2}\right].\label{eq26}
\end{equation}
or by solving for $k{{a}_{0}}$ we get
\begin{equation}
ka \simeq \sqrt{n(n^{2}-1)} \left(\frac{T}{\rho {{a}_{0}}v_z^2}\right)^{1/2}\left[1+\frac{3n-1}{4}\left(\frac{T}{\rho {{a}_{0}} v_z^2}\right)\right] \label{eq27}
\end{equation}
When the term proportional to $(k{{a}_{0}})^2$ is negligible, the equation is reduced to the wave number formula derived in the linear analysis, Eq.~(\ref{eq2}).
When $n=1$ (dipolar motion), Eq.~(\ref{eq27}) is satisfied if and only if $k=0$.
The dipolar motion corresponds to a constant translation of the center of cross-sectional mass.
It can be neglected as far as surface oscillation is concerned.
Therefore, the quadrupolar ($n=2$) motion is the lowest oscillatory motion.

Besides the Laplace equation, the second ordered equations are more complicated.
\begin{eqnarray}
&-&\frac{T}{{\rho}{{a}_{0}}}\left({F}^{(2)}+\frac{{\partial}^{2}{F}^{(2)}}{\partial{\theta}^{2}}+{{a}_{0}}^{2}\frac{{\partial}^{2} {F}^{(2)}}{\partial{z}^{2}}\right)+v_{z}\frac{\partial{\phi}^{(2)}}{\partial{z}}\nonumber\\
&+&\frac{T}{{\rho}{{a}_{0}}} \left[-2{\left(F^{(1)}\right)}^{2}+{\left(\frac{\partial{F}^{(1)}}{\partial\theta}\right)}^{2}-4{{a}_{0}}^{2}{F}^{(1)}\frac{{\partial}^{2}{F}^{(1)}}{\partial{z}^{2}} \right. \nonumber\\
& & \left.-{{a}_{0}}^{2}{\left(\frac{\partial{F}^{(1)}}{\partial{z}}\right)}^{2}\right] 
+ {\left(\frac{\partial{\phi}^{(1)}}{\partial{r}}\right)}^{2}
+\frac{1}{{{a}_{0}}^{2}}{\left(\frac{\partial{\phi}^{(1)}}{\partial\theta}\right)}^{2}\nonumber\\
&+&{\left(\frac{\partial{\phi}^{(1)}}{\partial{z}}\right)}^{2}
+4v_{z}{F}^{(1)}\frac{\partial{\phi}^{(1)}}{\partial{z}}+2{{a}_{0}}v_{z}{F}^{(1)}\frac{{\partial}^{2}{\phi}^{(1)}}{\partial{z}\partial{r}}=0, \nonumber\\\label{eq28}
\end{eqnarray}
\begin{eqnarray}
\frac{\partial{\phi}^{(2)}}{\partial{r}}&-&{{a}_{0}}v_{z}\frac{\partial{F}^{(2)}}{\partial{z}}
+4{F}^{(1)}\frac{\partial{\phi}^{(1)}}{\partial{r}}-\frac{2}{a}\frac{\partial{\phi}^{(1)}}{\partial\theta}\frac{\partial{F}^{(1)}}{\partial\theta} \nonumber\\
&-&4{{a}_{0}}v_{z}{F}^{(1)}\frac{\partial{F}^{(1)}}{\partial{z}} -2{{a}_{0}}\frac{\partial{\phi}^{(1)}}{\partial{z}}\frac{\partial{F}^{(1)}}{\partial{z}} \nonumber\\
&+&2{{a}_{0}}{F}^{(1)}\frac{{\partial}^{2}{\phi}^{(1)}}{\partial{r}^{2}}=0.
\label{eq29}
\end{eqnarray}
By substituting the expressions in Eqs.~(\ref{eq24-2}) and (\ref{eq27}) into ${F}^{(1)}$ and ${\phi}^{(1)}$, Eqs.~(\ref{eq28}) and (\ref{eq29}) become
\begin{eqnarray}
&-&\left({F}^{(2)}+\frac{{\partial}^{2}{F}^{(2)}}{\partial{\theta}^{2}}+a^{2}\frac{{\partial}^{2} {F}^{(2)}}{\partial{z}^{2}}\right) \nonumber\\
&+&\frac{T}{\rho{{a}_{0}}}\left[ \frac{{(n^{2}-1)}^{2}}{{v_{z}}^{2}}+\frac{2(n^{2}-1){{a}_{0}}^{2}k^{2}}{{v_{z}}^{2}}\right]{\cos}^{2}n\theta{\cos}^{2}kz \nonumber\\
&+&\frac{\rho{{a}_{0}}v_{z}}{T}\frac{\partial{\phi}^{(2)}}{\partial{z}}+\left[n^{2}{\sin}^{2}n\theta{\cos}^{2}kz \right.\nonumber\\
& &-\left.(2-4{{a}_{0}}^{2}k^{2}){\cos}^{2}n\theta{\cos}^{2}kz-{{a}_{0}}^{2}k^{2}{\cos}^{2}n\theta{\sin}^{2}kz\right]\nonumber\\
&+&\frac{\rho{{a}_{0}}}{T}{{a}_{0}}^{2}k^{2}{v_{z}}^{2}{\cos}^{2}n\theta{\sin}^{2}kz+\frac{T}{\rho{{a}_{0}}}\left[\frac{n^{2}{(n^{2}-1)}^{2}}{{{a}_{0}}^{2}k^{2}{v_{z}}^{2}}\right.\nonumber\\
& &+\left.\frac{2n^{2}(n^{2}-1)}{{v_{z}}^{2}}+\frac{n^{2}{{a}_{0}}^{2}k^{2}}{{v_{z}}^{2}}\right]{\sin}^{2}n\theta{\sin}^{2}kz\nonumber\\
&-&4(n^{2}-1+{{a}_{0}}^{2}k^{2}){\cos}^{2}n\theta{\cos}^{2}kz\nonumber\\
&-&\frac{2\rho{{a}_{0}}}{T}{{a}_{0}}^{2}k^{2}{v_{z}}^{2}{\cos}^{2}n\theta{\cos}^{2}kz=0, \label{eq30}
\end{eqnarray}
\begin{eqnarray}
& &\frac{\partial{\phi}^{(2)}}{\partial{r}}-{{a}_{0}}v_{z}\frac{\partial{F}^{(2)}}{\partial{z}}\nonumber\\
&+&\frac{T}{\rho{{a}_{0}}}\frac{1}{{{a}_{0}}kv_{z}}n^{2}(n^{2}-1+{{a}_{0}}^{2}k^{2}){\sin}^{2}n\theta\sin2kz\nonumber\\
&-&\frac{T}{\rho{{a}_{0}}}\frac{{{a}_{0}}k}{v_{z}}n^{2}(n^{2}-1+{{a}_{0}}^{2}k^{2}){\cos}^{2}n\theta\sin2kz=0. \nonumber\\
\label{eq31}
\end{eqnarray}
The terms comparable to or smaller than ${{a}_{0}}^{4}k^{4} T/(\rho{{a}_{0}})$ have been neglected since the sum of them would contribute the final result by no more than 0.01\%.

Now we are ready to evaluate the second ordered motion $F^{(2)}$.
The velocity potential must satisfies both Laplace equation and Eq.~(\ref{eq31}).  
Therefore, $\phi^{(2)}$ and $F^{(2)}$ can be written as

\begin{eqnarray}
\phi^{(2)}&=&J_{0}(i2kr)\gamma_{0}(z)+J_{2n}(i2kr)\cos2n\theta\gamma_{2n}(z), \nonumber\\
F^{(2)}&=&\delta_{0}(z)+\cos2n\theta\delta_{2n}(z).
\label{eq32}
\end{eqnarray}

The function $\delta_{0}(z)=\sin^2 kz$ is related to the area conservation.
Carrying on the calculation further yields the expression for $\delta_{2n}(z)$.
\begin{eqnarray}
\delta_{2n}(z)&=&\frac{2n^{3}+5n^{2}-2n-2}{4(4n^{2}-1)}+\frac{n^{2}+n+1}{4n(4n^{2}-1)}{{a}_{0}}^{2}k^{2}\nonumber\\
&+&\left[\frac{-2n^{3}+7n^{2}+2n-4}{4(2n^{2}+1)}\right.\nonumber\\
&+&\left.\frac{-40n^{5}+22n^{4}+23n^{3}+17n^{2}+11n+1}{4n(2n+1){(2n^{2}+1)}^{2}}{{a}_{0}}^2k^2\right] \nonumber\\
&\times& \cos2kz .
\label{eq33}
\end{eqnarray}
The expression for $B_{n=2}$, the coefficient attached to $\eta^2 \cos 4\theta$, is then obtained from $\delta_{4}(z)$ with $n=2$ (octapole component) and $\cos 2kz=1$ (nodal and anti-nodal positions):
\begin{eqnarray}
B_{n=2}&=&\frac{1}{2}\left.\delta_4(z)\right|_{\cos2kz=1}=\frac{5}{12}-\frac{232}{3240}{{a}_{0}}^{2}k^{2}\nonumber\\
&\simeq& 0.417-0.07{{a}_{0}}^{2}k^{2}. \label{eq34'}
\end{eqnarray}
which is Eq.\ (\ref{eq34}) quoted in Sec.\ \ref{correction}.


\begin{references}
\bibitem{note2}
Since Bohr's intention was to measure the surface tension, 
no attention was paid to the boundary profile at extreme positions by himself. 
Likewise, the aspect of his work on the cross-sectional boundary profile has been completely unrecognized.
\end{references}
\end{document}